\documentclass{jetpl}
	\usepackage[english]{babel}
	\usepackage{graphics}
	\usepackage[colorlinks=true, citecolor=blue, urlcolor=blue, linkcolor=black]{hyperref}
	\usepackage{cite}
	\usepackage{caption}

\twocolumn

\title{High quality factor mechanical resonance in a silicon nanowire}

\rtitle{High quality factor mechanical resonance in a silicon nanowire}

\sodtitle{High quality factor mechanical resonance in a silicon nanowire}

\author{D.\,E.~Presnov$^{\dag,\circ}$, S.~Kafanov$^\ddag$, A.\,A.~Dorofeev$^{\dag}$, I.\,V.~Bozhev$^{\dag}$, A.\,S.~Trifonov$^{\dag,\circ}$, Yu.\,A.~Pashkin$^{\ddag,\star}$\thanks[1]{\href{mailto: y.pashkin@lancaster.ac.uk}{y.pashkin@lancaster.ac.uk}}  V.\,A.~Krupenin$^{\dag}$\thanks[2]{\href{mailto: krupenin@physics.msu.ru}{krupenin@physics.msu.ru}}}

\rauthor{Presnov, Kafanov, Dorofeev, Bozhev, Trifonov, Pashkin, Krupenin}

\sodauthor{Presnov, Kafanov, Dorofeev, Bozhev, Trifonov, Pashkin, Krupenin}

\address{
	$^\dag$Quantum Technology Centre, Faculty of Physics, M.\,V. Lomonosov Moscow State University, Moscow, 119991, Russia.\\
	$^\circ$D.\,V. Skobeltsyn Institute of Nuclear Physics, M.\,V. Lomonosov Moscow State University, Moscow, 119991, Russia.\\
	$^\ddag$Department of Physics, Lancaster University, Lancaster, LA1~4YB, United Kingdom.\\
	$^\star$Lebedev Physical Institute, Moscow 119991, Russia.\\
	}

\dates{\today}{*}

\abstract{
	Resonance properties of nanomechanical resonators based on doubly clamped silicon nanowires, fabricated from silicon-on-insulator and coated with a thin layer of aluminum, were experimentally investigated. Resonance frequencies of the fundamental mode were measured at a temperature of $20\,\mathrm{mK}$ for nanowires of various sizes using the magnetomotive scheme. The measured values of the resonance frequency agree with the estimates obtained from the Euler-Bernoulli theory. The measured internal quality factor of the $5\,\mathrm{\mu m}$-long resonator, $3.62\times10^4$, exceeds the corresponding values of similar resonators investigated at higher temperatures. The structures presented can be used as mass sensors with an expected sensitivity $\sim 6 \times 10^{-20}\,\mathrm{g}\,\mathrm{Hz}^{-1/2}$.}


\begin{document}
	\renewcommand\figurename{Fig.}
	\maketitle

\section*{Introduction}
	 Nanoelectromechanical systems have been widely used in the fundamental and applied physics research \cite{RSI_2005, Zang_2015, Arash_2015, Pashkin_2012}, due to the rapid development of technology in recent years. Various microelectromechanical systems have become an integral part of our daily lives. Such structures are present almost everywhere, be it a phone or a car, in the form of various micron-scale accelerometers, gyroscopes, \textit{etc.} Today, when electronic devices make use of the properties of individual atoms and molecules \cite{Shorokhov_2017, Lovat_2017, Soldatov_1998_UFN}, applications of nanoelectromechanical systems include ultrasensitive detection of mass, down to the mass of single molecules \cite{Bartsch_2014, Yang_2004}, force \cite{Mamin_2001}, pressure \cite{Zhao_2012} and displacement \cite{Knobel_2003, Shevyrin_2015}. By coupling nanomechanical resonators with optical and electronic transducers, it became possible to explore various quantum effects \cite{Naik_2006, Teufel_2011, Harrabi_2012}. Also, small-size mechanical resonators are effective tools in understanding properties of superfluids \cite{Bradley_2017}.
	
	The operation principle of nanoelectromechanical mass sensors is based on the phenomenon of mechanical resonance that occurs when oscillations are excited, for example, in beams clamped at both ends, with a sub-micron cross-sectional area. The eigenfrequency of the resonator depends on its geometry and material, and increases when the resonator dimensions get smaller. It was shown \cite{Carr_1999} that the quality factor of resonators depends on their surface-to-volume ratio, and is also affected by the surface quality of the fabricated structure. The change of the resonator mass leads to the shift of its resonance frequency. The minimum detectable change of the mass is given by the expression \cite{Ekinci_2004}:
	\begin{equation}\label{eq1}
		\delta m=-2 \frac{0.735m}{f_0} \delta f_0,
	\end{equation}
	where  $m$ is the initial nanowire mass, $\delta m$ is the change of the mass, $f_0$ is the resonance frequency, and $\delta f_0$ is the minimum detectable frequency shift determined by the internal noise of the resonator and noise of the measurement system. Thus, to build an ultra-sensitive mass sensor, it is desirable to reduce the effective mass of the resonator and increase its resonance frequency, while simultaneously increasing its quality factor in order to resolve as small frequency shift as possible. Up to now, the record value reported for the quality factor of silicon resonators with a sub-micron cross-section was $\sim1.8\times 10^4$ \cite{Cleland_1996}.
	
	In the past decade, silicon-on-insulator (SOI) became a commonly used material for the fabrication of silicon nanoelectromechanical systems \cite{Mori_2014}. It comes as a three-layer wafer, in which the upper thin layer of single-crystal silicon is separated from the base substrate by a thin inter-layer of the silicon oxide. This material is used for the fabrication of field-effect transistors with a nanowire channel \cite{Presnov_2012, Presnov_2013}, which became the basis of biosensors for detecting molecules and viruses \cite{Rubtsova_2017}, as well as for nanoscale field charge sensors of an atomic force microscope for monitoring charge dynamics in various structures \cite{Trifonov_2017}.
	
	In this Letter, we briefly describe a technology for fabricating nanomechanical resonators from silicon-on-insulator based on the standard microelectronics processes, only. The main parameters of the resonators, \textit{viz.}, its eigenfrequencies and quality factors were determined experimentally. The dynamics of resonators in the linear and nonlinear regimes was investigated.

\section*{ Oscillation properties of nanowires}
	The design of the most common types of nanomechanical resonators is based on a suspended beam clamped at one or both ends. Such resonators have different vibration modes of bending, twisting, deformation, \textit{etc.} The bending mode is the most interesting, since it produces the maximum response to an external drive. This mode is easier to excite and detect by converting mechanical vibrations into an electrical signal. The oscillation dynamics of a suspended doubly clamped beam is generally described by the Euler-Bernoulli theory and well approximated by the motion of a simple harmonic oscillator with small damping \cite{Cleland_Book}. For the fundamental mode, the equation of motion of the midpoint of the beam excited by an external force has the following form:
	\begin{equation}\label{res}
		\ddot{x} + \gamma \dot{x} + \omega_0^2 x = f(t),
	\end{equation}
	where $x$ is the displacement of the beam midpoint from the equilibrium position (in the absence of an external force), $\gamma$ is the resonator damping constant, $\omega_0 = 2\pi f_0$ is the angular resonance frequency, and $f(t)$ is the external force.

	For oscillations parallel to the plane of the substrate, the resonance frequency of the fundamental bending mode is expressed by the following expression \cite{Cleland_Book}:
	\begin{equation}\label{eq2}
		f_0 = 1.03 \sqrt{\frac{E}{\rho}}\frac{w}{L^2},
	\end{equation}
	where $E$ and $\rho$ are the Young's modulus and density of the substrate material, respectively, and $w$ and $L$ are the width and length of the suspended nanowire. This equation is valid in the absence of the nanowire tension, which in principle can arise when the sample is cooled down because of the difference of the coefficients of thermal expansion of the materials used.

	The resonator remains in the linear regime at small mechanical displacements, however, at a higher driving force, the resonator can enter the nonlinear regime, which is accounted for by adding an extra term $\propto x^3$ to the left-hand side of Eq.\,(\ref{res}). In the nonlinear regime, the response curve becomes asymmetric \cite{Nayfeh_Mook_Book, Postma_2005, Tajaddodianfar_2017, Laurent_2017}. Depending on the configuration of the resonant system, the resonance frequency may increase (the resonator becomes ``harder'') or decrease (the resonator becomes ``softer''), which is taken into account by the sign of the cubic term. The resonator becomes a nonlinear system at oscillation amplitudes exceeding the critical amplitude $a_c$ expressed as \cite{Nayfeh_Mook_Book}:
	\begin{equation}\label{ac}
		a_c = 2 f_0\frac{L^2}{\pi} \sqrt{\frac{\rho\sqrt{3}}{EQ}},
	\end{equation}
	where $Q$ is the quality factor of the resonator.

	To study such resonators, the magnetomotive method is commonly employed \cite{Cleland_1996, Li_2008}: the resonator is placed in a constant uniform magnetic field directed perpendicular to the main axis of the nanowire (to the substrate) and an RF current is passed through the nanowire. As a result, under the Lorentz force, the nanowire starts to bend perpendicular to the direction of the magnetic field and the direction of the current flow. The resonance frequencies of the system can be found from the frequency dependence of the transmission coefficient of the RF signal. Only odd modes can be detected by this method since the induced emf vanishes for even modes due to the symmetry of the system. The fundamental bending mode corresponding to the half of the oscillation wavelength has the maximum displacement amplitude hence the largest induced response.

\section*{Sample fabrication}
	\begin{figure}[b]
		\includegraphics[width=\linewidth]{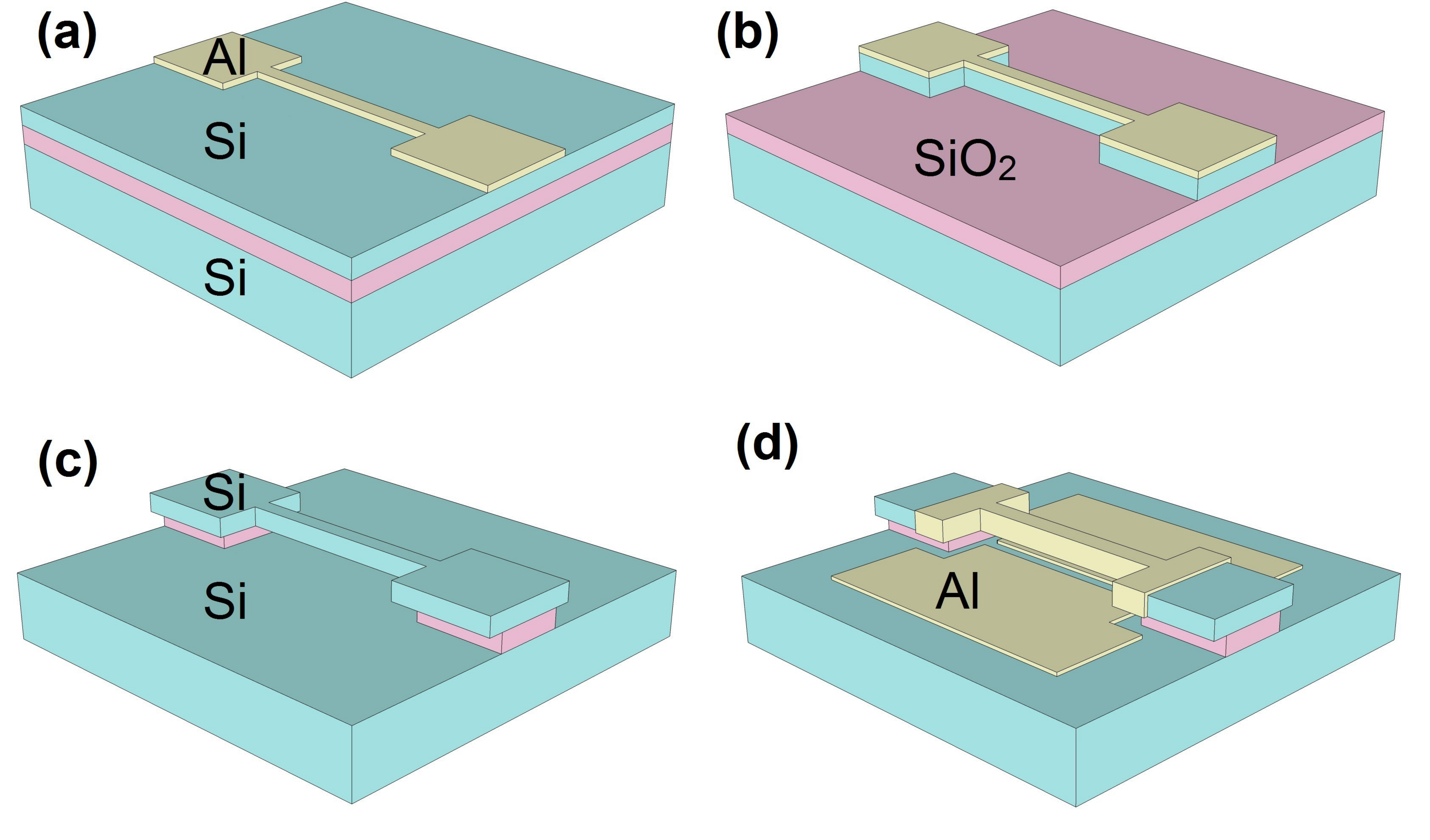}
		\caption{
			(color on-line) Fabrication stages of a silicon nanomechanical resonator:
			(a) aluminum mask formed on top of the SOI wafer;
			(b) nanowire structure in the top silicon layer after reactive ion etching;
			(c) suspended structure of the nanowire after removal of the silicon oxide layer;
			(d) final structure after metallization with a $20\,\mathrm{nm}$-thick layer of aluminum.
		}
		\label{Fig1}
	\end{figure}
	Conventional SOI wafers with a thickness of the upper silicon layer of $110\,\mathrm{nm}$, separated from the silicon substrate by a layer of the $200\,\mathrm{nm}$-thick silicon oxide were used for the sample fabrication. The fabrication process is similar to that used for sensors based on the field-effect transistors with a nanowire channel \cite{Presnov_2013,  MicroEl, Trifonov_2017} and involved three stages of electron-beam lithography with a positive resist and also reactive ion and wet etchings. The main fabrication stages are shown in Fig.\,\ref{Fig1}.

\begin{figure}[t]
		\centering
		\includegraphics[width=0.9\linewidth]{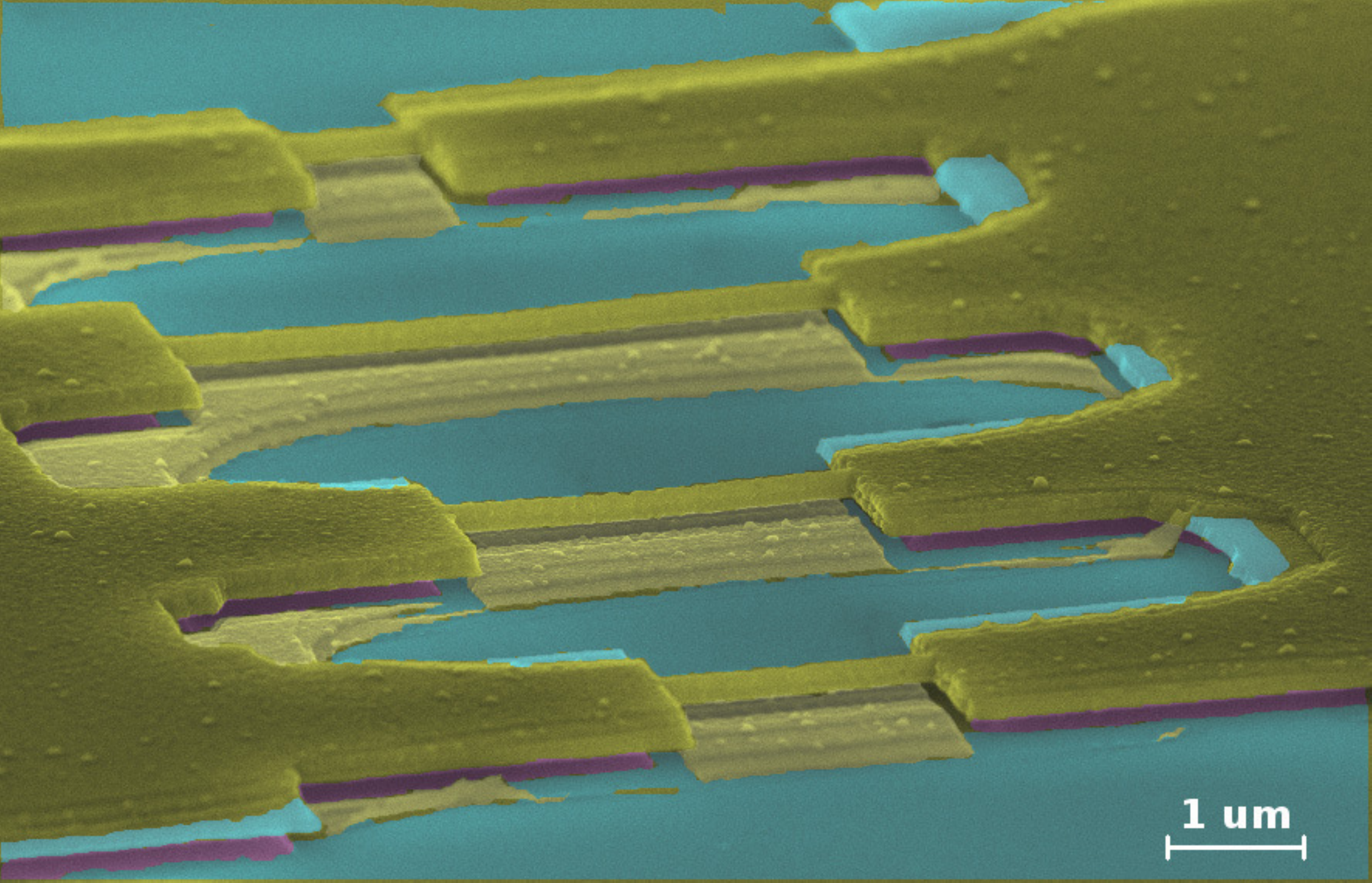}
		\caption{(Color online) False color image of a structure of four nanowires in a scanning electron microscope. Blue color denotes the silicon support substrate; magenta -- an insulating layer of silicon oxide; light-yellow -- silicon nanowires and wiring electrodes covered with a layer of aluminum, as well as parasitic shadows on the substrate formed after aluminum deposition at three different angles. The estimated resonators dimensions are (length $\times$ width) $5\,\mathrm{\mu m} \times 115\,\mathrm{nm}$, $3\,\mathrm{\mu m} \times 90\,\mathrm{nm}$, $2\,\mathrm{\mu m} \times 85\,\mathrm{nm}$, $1\,\mathrm{\mu m} \times 80\,\mathrm{nm}$.}
		\label{Fig2}
	\end{figure}

	Four resonant structures of different lengths connected in parallel were formed on the chip. With this layout, it is possible to measure all resonators in one cooldown, since the resonators do not affect each other due to the significant difference in their resonance frequencies. The disadvantage of this layout is that the incoming RF current is distributed among all resonators, which does not allow to estimate of the absolute value of the magnitude of the external force. A scanning electron microscope image of the final structure with the designed lengths of $1$, $2$, $3$ and $5\,\mathrm{\mu m}$, is shown in Fig.\,\ref{Fig2}.

\section*{Experimental results and discussion}
\begin{figure}[t]
		\includegraphics[width=0.97\linewidth]{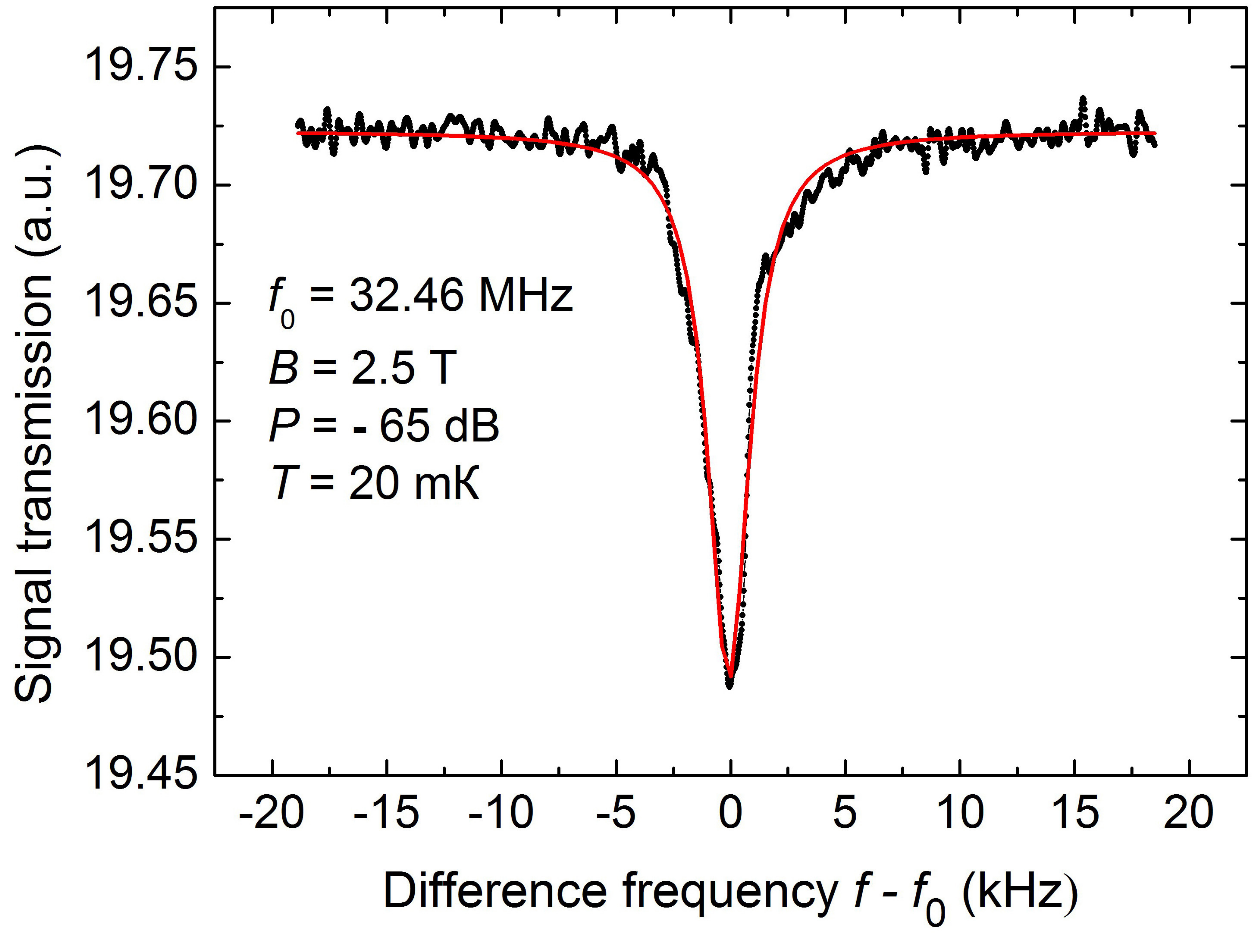}
		\caption{(Color online) Dependence of the signal transmission coefficient of the RF signal on the frequency for a suspended $5\,\mathrm{\mu m}$ nanowire (black) and the approximation of experimental data by the Lorentz function (red).}
		\label{Fig3}
	\end{figure}

	Resonance characteristics of the fabricated structures were investigated in a vacuum, in a magnetic field up to $5\,\mathrm{T}$ at a temperature of $20\,\mathrm{mK}$. The used frequency range was up to $500\,\mathrm{MHz}$ with a $200\,\mathrm{Hz}$ measurement bandwidth. RF signal from the network vector analyzer was attenuated by $35\,\mathrm{dB}$ at cryogenic temperatures and applied to the structure. The transmitted signal was amplified by $40\,\mathrm{dB}$ at room temperature and detected. The power values in the article correspond to the power applied to the measured structure. The shape of the frequency characteristic of the signal transmission coefficient is caused by the negative emf during the motion of a conductor in a magnetic field, has a characteristic minimum at the resonant frequency as it shown in Fig.\,\ref{Fig3}. The symmetry of this dependence indicates the linear regime of the resonator under the small external driving force. By approximating the experimental data by the Lorentz function, the resonance frequency and $Q$-factor of the resonators are determined. The measured values of the resonance frequencies for the $5$, $3$ and $2\,\mathrm{\mu m}$-long nanowires were found to be $32.46$, $71.99$ and $150.25\,\mathrm{MHz}$, respectively. This agrees within the 2\% error with the theoretical estimates obtained from Eq.\,(\ref{eq2}). The estimated value of the resonance frequency for the $1\,\mathrm{\mu m}$-long nanowire is outside of the measured frequency range.
	
	The mass of the resonator $\sim 1.2 \times 10^{-13}\,\mathrm{g}$ is estimated from its geometric dimensions. The minimum detectable frequency noise $\sim8\,\mathrm{Hz}\,\mathrm{Hz}^{-1/2}$ is determined from the magnitude of the signal transmission coefficient fluctuations $10^{-3}\,\mathrm{a.\,u.}\,\mathrm{Hz}^{-1/2}$ and the greatest value of the response gradient $125 \times 10^{-6}\,\mathrm{a.\,u.}\,\mathrm{Hz^{-1}}$. By plugging these numbers into Eq.\,(\ref{eq1}) one can estimate the mass sensitivity in the linear regime $\sim6 \times 10^{-20}\,\mathrm{g}\,\mathrm{Hz}^{-1/2}$, which is comparable with the values of mass sensitivity obtained by other groups \cite{Ekinci_2004}. The minimum detectable mass is $\sim 8.5 \times 10^{-19}\,\mathrm{g}$, which corresponds, for example, to a silicon sphere of radius $4.4\,\mathrm{nm}$. When the sensor operates in the nonlinear regime, its mass sensitivity can be significantly improved \cite{Buks_2006}.
	
	\begin{figure}[t]
		\includegraphics[width=\linewidth]{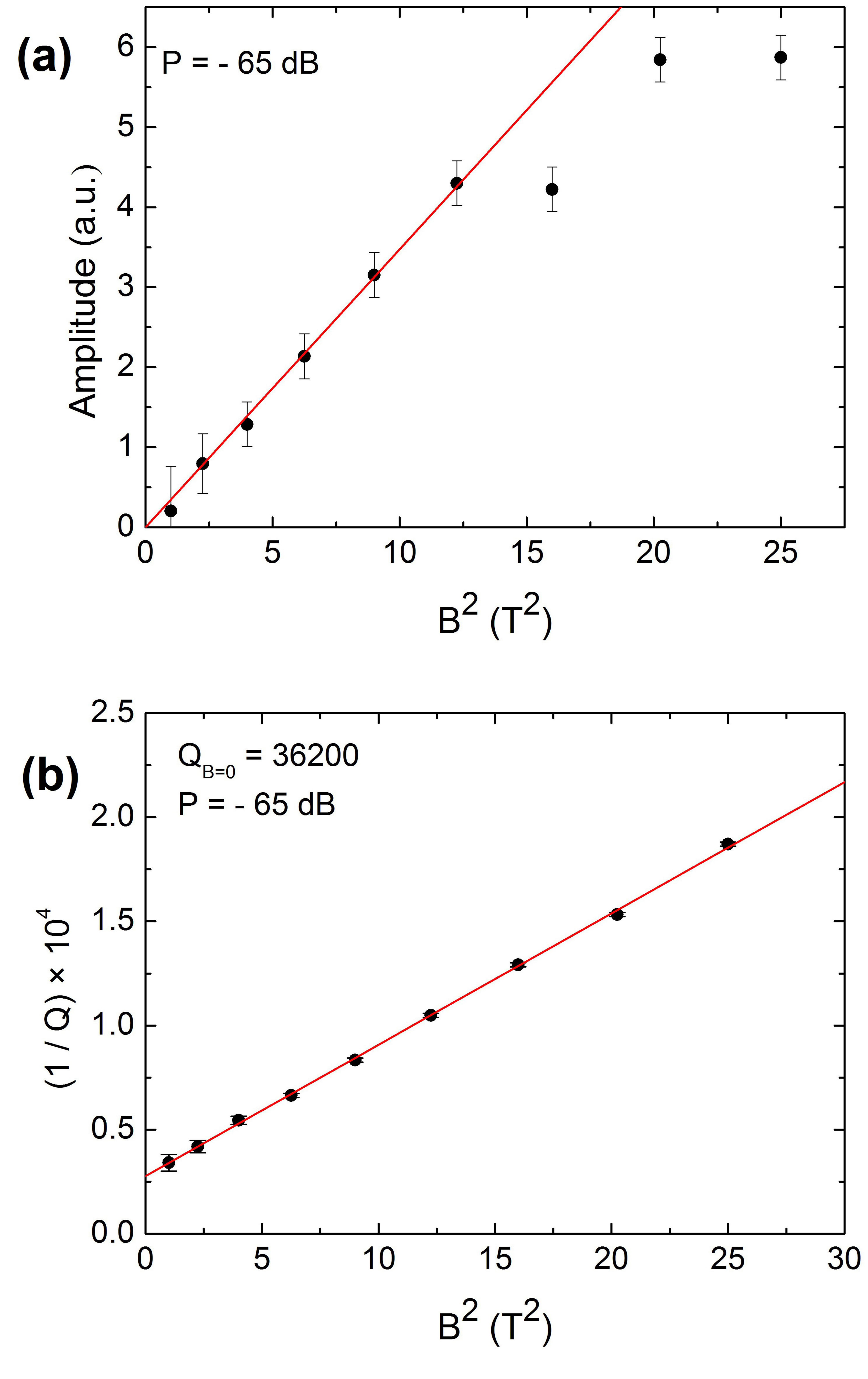}
		\caption{(a) Dependence of the oscillation amplitude of $5\,\mathrm{\mu m}$ on the square of the magnetic field strength and its linear approximation (straight line). (b) Dependence of the inverse $Q$-factor values of the resonator on the square of the magnetic field strength and its linear approximation.}
		\label{Fig4}
	\end{figure}
	As shown in \cite{Cleland_1999}, when a mechanical resonator is excited by a magnetomotive method in the linear regime, the displacement amplitude is proportional to the square of the magnetic field and to the amplitude of the RF current, $I\propto P^{1/2}$, where $P$ is the RF power. Figure\,\ref{Fig4}(a) shows the dependence of the oscillation amplitude of the $5\,\mathrm{\mu m}$-long resonator on the square of the magnetic field $B$. In magnetic fields up to $3.5\,\mathrm{T}$ the experimental data are well approximated by the expected linear dependence of the amplitude of mechanical oscillations on the magnitude of the magnetic induction. With an increase in the magnetic field above $3.5\,\mathrm{T}$, the resonator begins to enter the nonlinear regime and the amplitude increase becomes weaker.

	The width of the resonant response $\Delta f$ is determined by the losses both in the resonator itself -- internal losses, and in the measurement system -- external losses. For strong magnetic fields, the main ones are magnetomotive losses which are proportional to the square of the magnetic field, $Q^{-1} \propto \Delta f \propto B^2$. The measured quality factor increases from $6.0 \times 10^3$ at $5\,\mathrm{T}$ to $2.95 \times 10^4$ at $0.5\,\mathrm{T}$. The experimental dependence of $Q^{-1}$ on the square of the magnitude of the magnetic field is shown in Fig.\,\ref{Fig4}(b). The linearity of this dependence indicates the dominance of magnetomotive losses over internal ones. Such dependence is preserved throughout the investigated range of magnetic fields. The internal $Q$-factor of the resonator is estimated to be $3.62 \times 10^4$ by the extrapolation to zero magnetic field. Such $Q$-factor values exceed those obtained earlier in similar structures made from SOI \cite{Yu_2012} and are comparable to the values in structures made of silicon \cite{PRB72, NL}, silicon coated with aluminum \cite{PRB81} and diamond \cite{PRB79}. It was confirmed in the earlier experiments \cite{PRB72, PRB81, PRB79} that losses in resonators (which are proportional to the inverse $Q$-factor values) decrease at temperatures below $1\,\mathrm{K}$, but the nature of this dependence has not been understood yet. The development of a reliable physical model describing losses in such systems at low temperatures is at present the subject of scientific research. In our experiments, measurements of the temperature dependence of the $Q$-factor were not carried out, but it is obvious that the low temperature is crucial for obtaining the high $Q$-factor of nanomechanical resonators. 
	
	\begin{figure}[t]
		\centering
		\includegraphics[width=0.99\linewidth]{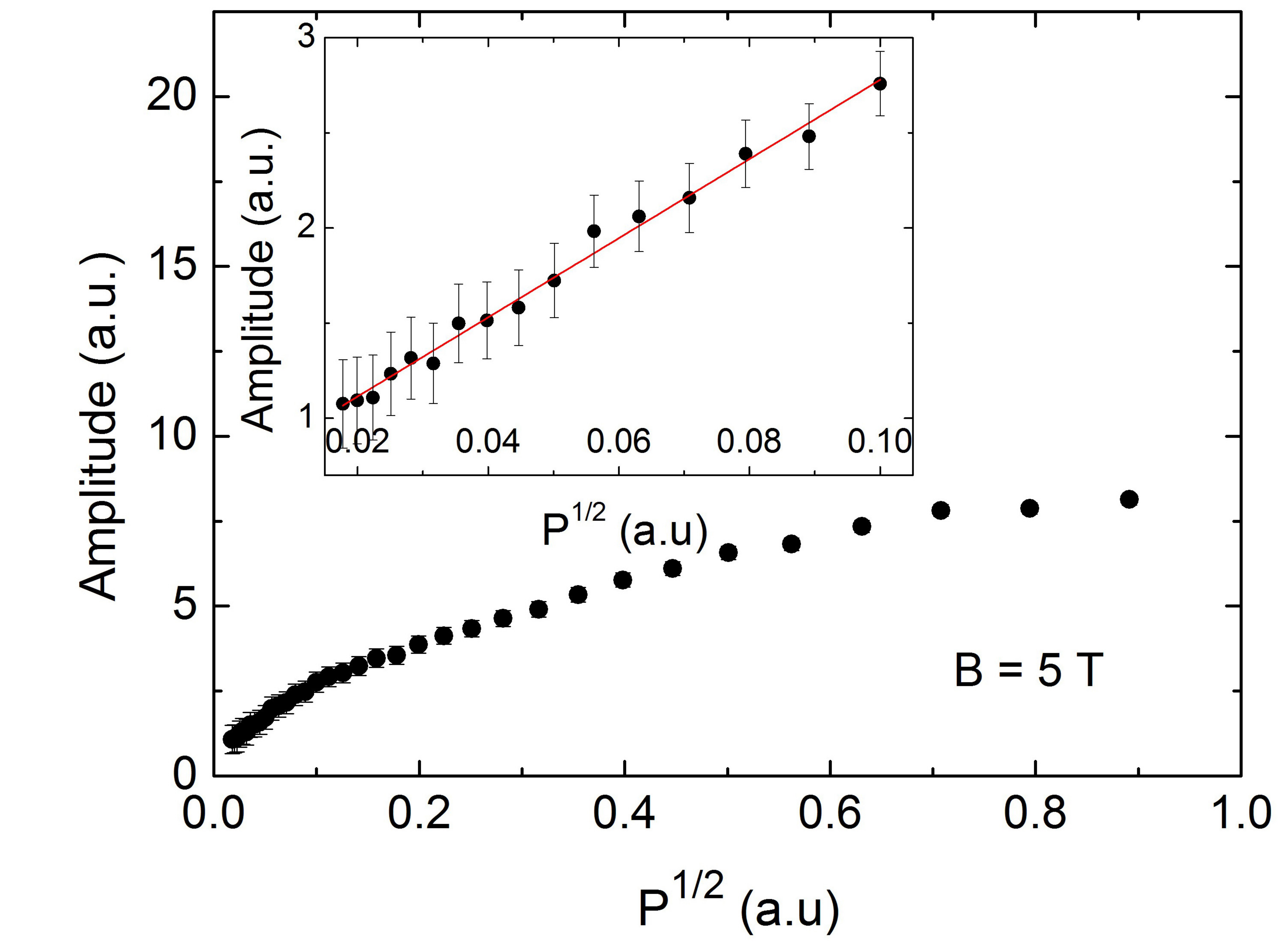}
		\caption{Dependence of the oscillation amplitude of the $2\,\mathrm{\mu m}$-long resonator on the square root of the input power. The inset shows the linear part of the entire dependence at low RF signal powers up to $0.1\,\mathrm{a.\,u.}$. Straight line is the linear approximation of experimental data.}
		\label{Fig5}
	\end{figure}
\begin{figure}[t]
		\includegraphics[width=\linewidth]{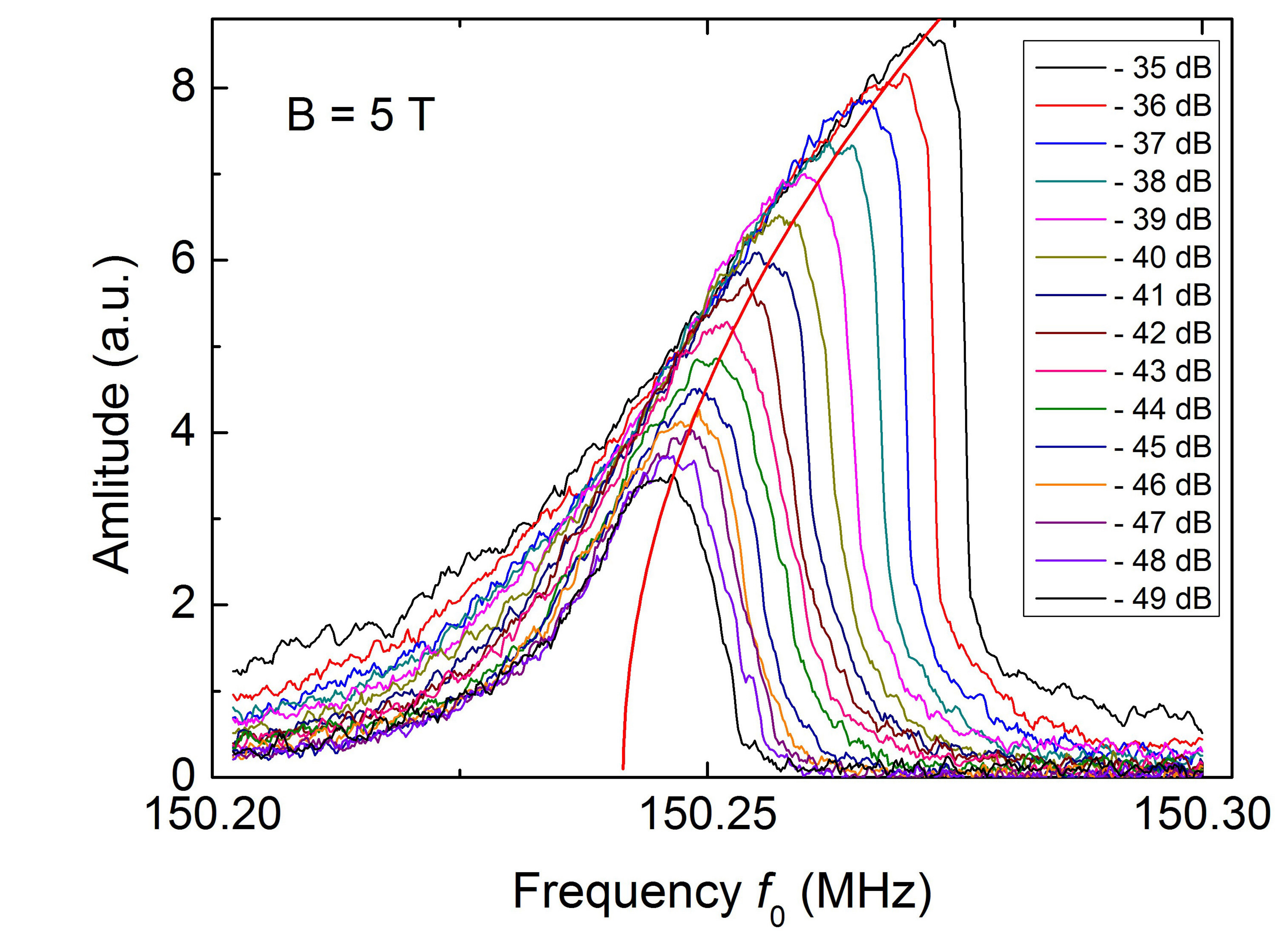}
		\caption{(Color online) A family of response curves of the $2\,\mathrm{\mu m}$-long resonator at high drive powers. When the RF power is increased, the resonator becomes ``harder''. The resonance frequency increases with the amplitude of the resonator oscillations. The position of the maxima of the resonance curves is approximated by a parabola, ``backbone curve'', from which the frequency of oscillations of the resonator in the linear regime is determined.}
		\label{Fig6}
	\end{figure}

	Figure\,\ref{Fig5} shows the dependence of the amplitude of mechanical oscillations on the square root of the power of the applied RF signal. The inset illustrates the fact that at low powers the amplitude of the oscillations is proportional to the oscillating force and the dynamics of the resonator is described by the linear equation of the harmonic oscillator, Eq.\,(\ref{res}). A slower increase in the amplitude of the oscillations with increasing driving force ($P^{1/2} > 0.1\,\mathrm{a.\,u.}$) indicates that the resonator becomes ``harder''. This also affects the shape of the response curve, which bends  towards the higher frequencies, as shown in Fig.\,\ref{Fig6}. It is easy to show that adding a cubic term $\propto x^3$ to Eq.\,(\ref{res}) leads to a quadratic dependence of the resonance frequency on the amplitude of the mechanical displacement \cite{Nayfeh_Mook_Book}, which is observed in our experiments. The transition to the nonlinear regime corresponds to the critical amplitude $a_c \sim 0.9\,\mathrm{nm}$ calculated from Eq.\,(\ref{ac}). From this, the largest amplitude of the  resonator oscillations $\sim 2.5\,\mathrm{nm}$ at the maximum pumping power of \cite{Postma_2005} is found.

\section*{ Conclusion }
	This Letter reports the resonance properties of the doubly clamped silicon nanowires. By using the magnetomotive measurements scheme, we determined the resonance frequencies of the fundamental bending modes of the nanowires in the range of $30-150\,\mathrm{MHz}$. The intrinsic $Q$-factor of the $5\,\mathrm{\mu m}$-long resonator was found to be $3.62 \times 10^4$, which is record high for resonators of similar dimensions fabricated from silicon-on-insulator. The measured values of the resonance frequencies are in good agreement with the estimates obtained from the Euler-Bernoulli theory. The dynamics of the resonators is described qualitatively by the model of the simple harmonic oscillator with a small dissipation and nonlinearity under the external driving force. The fabrication method for resonant structures uses standard silicon technological recipes only and can be used to fabricate high-sensitivity mass, force and displacement sensors. The mass sensitivity of the suspended $5\,\mathrm{\mu m}$-long nanowire is estimated to be $\sim 6 \times 10^{-20}\,\mathrm{g}\,\mathrm{Hz}^{-1/2}$.

	The work was supported by the Russian Foundation for Basic Research, RFBR Grant No. 16-29-03266 and the Royal Society (grant IEC\textbackslash R3\textbackslash170029). The equipment of the ``Educational Center of Lithography and Microscopy,'' M.\,V. Lomonosov Moscow State University was used.


\begin{thebibliography}{33}
	\bibitem{RSI_2005}
	Ekinci K.\,L., and Roukes M.\,L.,	\href{https://doi.org/10.1063/1.1927327}{Rev. Sci. Inst. \textbf{76}, 061101, (2005)}.
	
	\bibitem{Zang_2015}
	Zhang W-M., \textit{et al.}, \href{https://doi.org/10.3390/s151026478}{Sensors \textbf{15}, 26478, (2015)}.
	
	\bibitem{Arash_2015}
	Arash B., Jiang J-W., and Rabczuk T., \href{https://doi.org/10.1063/1.4916728}{Appl. Phys. Rev. \textbf{2}, 021301, (2015)}.
	
	\bibitem{Pashkin_2012}
	Greenberg Ya.\,S., Pashkin Yu.\,A., and Il'ichev E., \href{https://doi.org/10.3367/UFNr.0182.201204c.0407}{Phys. Usp. \textbf{55} 382, (2012)}.
	
	\bibitem{Shorokhov_2017}
	Shorokhov V.\,V., \textit{et al.}, \href{https://doi.org/10.1039/C6NR07258E}{Nanoscale \textbf{9}, 613, (2017).}
	
	\bibitem{Lovat_2017}
	Lovat G., \textit{et al.}, \href{https://doi.org/10.1038/nnano.2017.156}{Nat. Nanotech. \textbf{12}, 1050, (2017).}
	
	\bibitem{Soldatov_1998_UFN}
	Soldatov E.\,S., \textit{et al.}, \href{http://dx.doi.org/10.1070/PU1998v041n02ABEH000364}{Phys. Usp. \textbf{41} 202, (1998)}
	
	\bibitem{Bartsch_2014}
	Bartsch S.\,T., Arp M., and Ionescu A.\,M., 	\href{https://doi.org/10.1109/JEDS.2013.2295246}{IEEE J. Elect. Dev. Soc. \textbf{2}, 8, (2014).}
	
	
	\bibitem{Yang_2004}
	Ilic B., Yang Y., and Craighead H.\,G., 	\href{https://doi.org/10.1063/1.1794378}{Appl. Phys. Lett. \textbf{85}, 2604, (2004).}
	
	
	\bibitem{Mamin_2001}
	Mamin, H. and Rugar, D., \href{https://doi.org/10.1063/1.1418256}{Appl. Phys. Lett. \textbf{79}, 3358, (2001).}
	
	
	\bibitem{Zhao_2012}
	Zhao X., \textit{et al.}, \href{https://doi.org/10.1364/OE.20.008535}{Optics Express \textbf{20}, 8535, (2012)}.
	
	
	\bibitem{Knobel_2003}
	Knobel R.\,G., and Cleland A.\,N., 	\href{https://doi.org/10.1038/nature01773}{Nature \textbf{424}, 291, (2003).}
	
	
	\bibitem{Shevyrin_2015}
	Shevyrin A.\,A., \textit{et~al.}, \href{https://doi.org/10.1063/1.4920932}{Appl. Phys. Lett. \textbf{106}, 183110, (2015).}
	
	
	\bibitem{Naik_2006}
	Naik A., \textit{et al.}, \href{http://dx.doi.org/10.1038/nature05027}{Nature \textbf{443}, 193, (2006)}.
	
	
	\bibitem{Teufel_2011}
	Teufel J., \textit{et al.},  \href{http://dx.doi.org/10.1038/nature10261}{Nature \textbf{475}, 359, (2011)}.
	
	\bibitem{Harrabi_2012}
	Harrabi K., \textit{et~al.}, \href{https://doi.org/10.1007/s00339-012-6981-8}{Appl. Phys. A \textbf{108}, 7, (2012)}
	
	
	\bibitem{Bradley_2017}
	Bradley D.\,I., \textit{et al.}, 	\href{https://doi.org/10.1038/s41598-017-04842-y}{Sci. Rep. \textbf{7}, 4876, (2017).}
	
	\bibitem{Carr_1999}
	Carr D.\,W., \textit{et al.}, \href{https://doi.org/10.1063/1.124554}{Appl. Phys. Lett. \textbf{75}, 920, (1999)}.
	
	
	\bibitem{Ekinci_2004}
	Ekinci K.\,L., Huang X.\,M.\,H., and Roukes M.\,L, \href{https://doi.org/10.1063/1.1755417}{Appl. Phys. Lett. \textbf{84}, 4469, (2004).}
	
	\bibitem{Cleland_1996}
	Cleland A.\,N., and Roukes M.\,L., \href{https://doi.org/10.1063/1.117548}{Appl. Phys. Lett. \textbf{69}, 2653, (1996).}
	
	\bibitem{Mori_2014}
	Mori K.,     \href{https://doi.org/10.1533/9780857099259.2.435}{Silicon-On-Insulator (SOI) Technology, Chapter \textbf{14}, 435, Woodhead Publishing, (2014)}.
	
	\bibitem{Presnov_2012}
	Presnov D.\,E., Amilonov S.\,V., and Krupenin V.\,A.,
	\href{https://doi.org/10.1134/S1063739712050034}{Russian microelectronics, \textbf{41}, 310, (2012)}
	
	\bibitem{Presnov_2013}
	Presnov D.\,E., \textit{et al.}, \href{https://doi.org/10.3762/bjnano.4.38}{	Beilstein J. Nanotechnol. \textbf{4}, 330, (2013).}
	
	\bibitem{Rubtsova_2017}
	Rubtsova M., \textit{et~al.}, 	\href{https://doi.org/10.1016/j.protcy.2017.04.099}{``Biosensors 2016'', Procedia Technology, \textbf{27}, 234, (2017).}
	
	\bibitem{Trifonov_2017}
	Trifonov A.\,S., \textit{et~al.}, 	\href{https://doi.org/10.1016/j.ultramic.2017.03.030}{Ultramicroscopy \textbf{179}, 33, (2017)}.
	
	
	\bibitem{Cleland_Book}
	Cleland A.\,N., \href{https://doi.org/10.1007/978-3-662-05287-7}{Foundations of nanomechanics, Springer, Berlin, (2003)}
	
	
	\bibitem{Nayfeh_Mook_Book}
	Nayfeh A.\,H. and Mook D.\,T., \href{https://doi.org/10.1002/9783527617586}{   Nonlinear Oscillations.,  WILEY--VCH Verlag GmbH \& Co. KGaA. (2007)}.
	
	\bibitem{Postma_2005}
	Postma H.\,W.\,Ch., \textit{et al.}, \href{https://doi.org/10.1063/1.1929098}{Appl. Phys. Lett. \textbf{86}, 223105, (2005)}.
	
	\bibitem{Tajaddodianfar_2017}
	Tajaddodianfar F., Yazdi M.\,R.\,H., and Pishkenari H.\,N., 	\href{https://doi.org/10.1007/s00542-016-2947-7}{Microsyst. Technol. \textbf{23}, 1913, (2017).}
	
	\bibitem{Laurent_2017}
	Laurent L., \textit{et al.}, \href{https://doi.org/10.1016/j.sna.2017.06.027}{Sens. and Act. A: Phys. \textbf{263}, 326, (2017).}
	
	\bibitem{Li_2008}
	Li T.\,F., \textit{et al.}, \href{https://doi.org/10.1063/1.2838749}{Appl. Phys. Lett. \textbf{92}, 043112, (2008)}
	
	
	\bibitem{MicroEl}
	Presnov D.\,E., Amitonov S.\,V. and Krupenin V.\,A., \href{https://doi.org/10.1134/S1063739712050034}{Russian Microelectronics \textbf{41}, 364, (2012).}
	
	\bibitem{Buks_2006}
	Buks E., and Yurke B., \href{https://doi.org/10.1103/PhysRevE.74.046619}{Phys. Rev. E \textbf{74}, 046619, (2017).}
	
	\bibitem{Cleland_1999}
	Cleland A.\,N., and Roukes M.\,L., \href{https://doi.org/10.1016/S0924-4247(98)00222-2}{Sens. and Act., \textbf{72}, 256, (1999).}
	
	\bibitem{Yu_2012}
	Yu L., \textit{et al.}, \href{https://doi.org/10.1109/TNANO.2012.2212028}{IEEE Trans. Nanotech.   \textbf{11}, 1093, (2012).}
	
	\bibitem{PRB72}
	Zolfagharkhani G., \textit{et al.}, \href{https://doi.org/10.1103/PhysRevB.72.224101}{Phys. Rev. B \textbf{72}, 224101, (2005).}		
	
	\bibitem{NL}
	Sulkko J., \textit{et al.}, \href{https://doi.org/10.1021/nl102771p}{Nano Lett. \textbf{10}, 4884, (2010).}	
	
	\bibitem{PRB81}
	Hoehne F., \textit{et al.}, \href{https://doi.org/10.1103/PhysRevB.81.184112}{Phys. Rev. B \textbf{81}, 184112, (2010)}.		
	
	\bibitem{PRB79}
	Imboden M., Mohanty P., \href{https://doi.org/10.1103/PhysRevB.79.125424}{Phys. Rev. B \textbf{79}, 125424, (2009)}.		
\end{thebibliography}
\end{document}